\documentclass{ar-1col}
\usepackage[numbers]{natbib}

\jname{Xxxx. Xxx. Xxx. Xxx.}
\jvol{67}
\jyear{2017}
\doi{10.1146/((please add article doi))}

\begin{document}

\markboth{Thielemann et al.}{Nucleosynthesis in Neutron Star Mergers}

\title{Neutron Star Mergers and Nucleosynthesis of Heavy Elements}

\author{F.-K. Thielemann$^{1,2}$, M. Eichler$^{3}$, I.V. Panov$^{4,5}$, and B. Wehmeyer${^6}$
\affil{$^1$Department of Physics, University of Basel, CH-4056 Basel, Switzerland; email: f-k.thielemann@unibas.ch}
\affil{$^2$GSI Helmholtzzentrum f\"ur Schwerionenforschung GmbH, D-64291 Darmstadt, Germany}
\affil{$^3$Institut f\"ur Kernphysik, Technische Universit\"at Darmstadt, D-64289 Darmstadt, Germany: email marius.eichler@theorie.ikp.physik.tu-darmstadt.de}
\affil{$^4$Institute for Theoretical and Experimental Physics of NRC Kurchatov Institute, Moscow, Russia; email: panov\_iv@itep.ru}
\affil{$^5$Sternberg Astronomical Institute, M.V. Lomonosov State University,
 Universitetskij pr., 13, 119234, Moscow, Russia}
\affil{$^6$Department of Physics, North Carolina State University, Raleigh, NC, 27695-8202, USA; email: bwehmey@ncsu.edu}}

\begin{abstract}
Neutron star mergers have been predicted since the 1970's, supported by the discovery of the binary pulsar and the observation of its orbital energy loss, consistent with General Relativity. They are considered as nucleosynthesis sites of the rapid neutron-capture process (r-process), being responsible for making about half of all heavy elements beyond Fe and being the only source of elements beyond Pb and Bi. Detailed nucleosynthesis calculations based on the decompression of neutron-star matter are consistent with solar r-process abundances of heavy nuclei. More recently neutron star mergers have also been identified with short duration Gamma-Ray Bursts via their IR afterglow, only explainable by the opacities of heavy (rather than only Fe-group) nuclei.
Two other observations support rare events like neutron star mergers as a dominant scenario for the production of the heaviest r-process nuclei: (a)The discrepancy between the latest admixtures of two long-lived radioactivities ($^{60}$Fe and 
$^{244}$Pu) found on earth seems to exclude the origin of the latter from core collapse supernovae. (b)The ratio of [Eu/Fe], with Eu being dominated by r-process contributions, shows a strong scatter in low metallicity stars up to [Fe/H]$<$-2, arguing for a strongly reduced occurrence rate in comparison to core-collapse supernovae. 
The high neutron densities in ejected matter permit a violent r-process, encountering fission cycling of the heaviest nuclei in regions far from (nuclear) stability. Uncertainties in nuclear properties, like nuclear masses, beta-decay half-lives, fission barriers and fission fragment distributions affect the detailed abundance distributions. The modeling of the astrophysical events depends also on the hydrodynamic treatment, i.e. SPH vs. grid calculations, Newtonian vs. GR approaches, the occurrence of a neutrino wind after the merger and before the emergence of a black hole, and finally the properties of black hole accretion disks. We will discuss the effect of both (nuclear and modelling) uncertainties and conclude that binary compact mergers are probably a or the dominant site of the production of r-process nuclei in our Galaxy.
A small caveat exists with respect to explaining the behavior of [Eu/Fe] at lowest metallicities and the question whether neutron star mergers can already contribute at such early times in galactic evolution.
\end{abstract}

\maketitle

\tableofcontents

\section{INTRODUCTION}
\label{sec:1}
Neutron stars, postulated shortly after the discovery of the neutron, were predicted as the final fate of massive stars, ending in supernova events \cite{baadezwicky34}. Their existence was proven
in the 1960's after the first observations of pulsars
\cite{hewish65}. We have by now an extensive knowledge of the distribution of neutron star masses and the underlying equation of state, \citep[e.g.][]{ozel12,hebeler13,oertel16}, with the most precise determinations existing for binary systems. Shortly after the discovery of the binary pulsar \cite{hulsetaylor75}, with an energy loss in agreement with General Relativity, it was found that 
this system would merge in $10^8$ years. This led to the prediction that neutron star or neutron star - black hole mergers would eject r-process nuclei \cite{lattimer74,lattimer76,symbalisty82}, followed up by a first detailed analysis of possible abundance distributions \cite{meyer88}. Later predictions included that such mergers would be accompanied by neutrino bursts and
gamma-ray bursts \cite{eichler89}. The very first and later more precise estimates of the mass ejection from neutron star 
mergers in Newtonian approximation followed \cite{davies94,ruffert96,rosswog99,rosswog00}, together with the very first detailed nucleosynthesis predictions \cite{freiburghaus99}. 

More recently, extensive investigations have been undertaken with respect to nucleosynthesis predictions \cite{panov04,panov08,goriely11,korobkin12,panov13,bauswein13,goriely13,hotokezaka13,rosswog14,wanajo14,just15,goriely14,perego14,eichler15,martin15,mendoza15,ramirez15,hotokezaka15,just16,radice16,roberts17},
with new approaches going beyond a Newtonian treatment, via conformally flat to fully relativistic treatments \citep[e.g.][]{oechslin02,oechslin04,oechslin07,bauswein13,hotokezaka13,wanajo14,sekiguchi15,sekiguchi16}, including also the role of 
magnetic fields \cite{kiuchi15}. Modern simulations do not only consider the composition of the dynamical ejecta, but also a neutrino wind composition (along the poles), where matter is ejected from the combined (initially rotationally stabilized) hot neutron star 
\citep[e.g.][]{dessart09,korobkin12,perego14,martin15,wanajo14,sekiguchi15,sekiguchi16,lehner16,roberts17}, up to the point of black hole formation \cite{metzger14}, if the maximum neutron stars mass is exceeded, and afterwards ejection of matter takes place
from (viscous) black hole accretion disks. The outflow of black hole accretion disks has been
investigated in detail in a number of studies, \citep[e.g.][]{surman06,surman08,surman14,just15,fernandez15,just16,wu16} and the effect of neutrino
conversion via matter-neutrino resonances has been analyzed with respect to a possible impact on nucleosynthesis \cite{malkus12,foucart15,malkus16,zhu16,frensel17}. For a good overview of all these components, including jet formation and ejection see \cite{hotokezaka15b}.
In parallel to neutron star mergers also neutron star - black hole mergers have been investigated \citep[e.g.][]{rosswog05,korobkin12,wanajo12,kyutoku13,foucart14,mennekens14,rosswog17}

Such nucleosynthesis predictions have been extensively tested with respect to nuclear uncertainties due to masses far from 
stability, beta decays, fission barriers, and fission fragment distributions \citep[e.g.][]{goriely13,goriely14,eichler15,mendoza15,mumpower16,wu16,shibagaki16}. The effect of the nuclear equation of state was investigated as well 
\citep[e.g.][]{oechslin04,oechslin07,bauswein13,bauswein14,just15,sekiguchi15,bauswein16}.

There exists extensive literature relating these events to short duration Gamma-Ray Bursts and/or macronovae/kilonovae as electromagnetic counterparts \citep[for recent literature see e.g.][]{lipacz97,tanvir13,tanaka13,kasen13,grossman14,rosswog14,metzger14,wanderman15,fryer15,hotokezaka16,barnes16,rosswog17,metzger17}. 
However, this issue was the central feature of last year's review by Fernandez \& Metzger \cite{fernandez16}, and we refer here to that article.  
Although these objects are also of major importance as strong sources for gravitational wave emission \citep{abbott16}, underpinning the importance of multi-messenger observations, we will essentially focus.in the present review on the ejected nucleosynthesis composition. The nucleosynthesis is constrained  by solar r-process abundances and whether they can be reproduced by compact object mergers,  by observations of low metallicity stars which are affected by the occurence frequency 
as a function of time during galactic evolution, and finally by information from individual events which relate to the observed light curve and spectra (here the nucleosynthetic composition connects via its effect on opacities to the electrodynamic signal). This review covers observational
constraints in section 2, the required thermodynamic conditions and neutron-richness of the ejecta in section 3, and a detailed discussion of nucleosynthesis results from compact object mergers in section 4. Finally, in section 5 we come back to 
issues in galactic evolution and whether compact binary mergers can match observations in the early Galaxy, before presenting
conclusions in section 6.

\begin{figure}[h]
\includegraphics[scale=0.5]{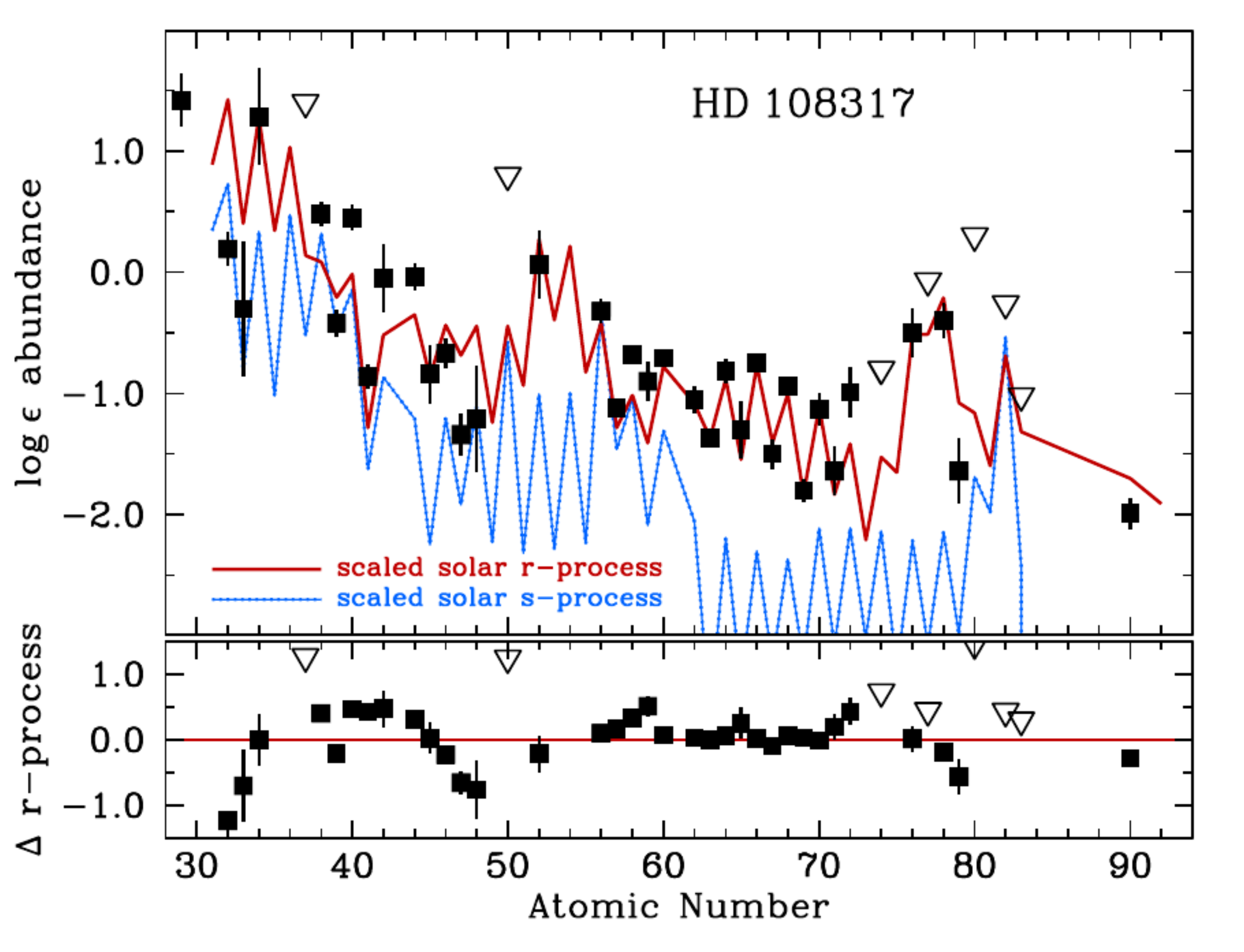}
%
%
\caption{Shown are the observed abundances of typical low metallicity stars which unveil a clear r-process (and not an s-process) pattern, exactly as found in the solar system, at least for elements with $Z\ge 40$ \citep{roederer14}.    }
\label{fig_1}       
\end{figure}

\section{OBSERVATIONAL EVIDENCE FOR r-PROCESS NUCLEOSYNTHESIS}
\label{sec:2}
\subsection{Solar r-Process Abundances and Patterns in Low-Metallicity Stars}

One interesting aspect to be tested relates to the question whether mergers lead to a quite robust r-process environment, 
which each time
produces the heavy r-elements (at least those with $A\ge130$) in proportions similar to solar \citep[see Fig.\ref{fig_1} and][]{sneden08,roederer12}. On the other hand  there exist variations in the contribution of lighter elements with $Z\le50$ \citep{qian07}. Could this be due to variations in the production site or require different production sites? A fraction of old metal-poor halo stars shows a large variety of abundance signatures,  including also r-elements like Eu \citep[see e.g.][] {honda06,hansen14,roederer17}. Possibly this 
is indicating a different weaker neutron-capture source, maybe a fraction of regular supernovae \cite{nishimura15,nishimura16}? Finally it should also be noted, that not in all low-metallicity
star observations Th and U show up in solar proportions (or with appropriate abundances due to their decay since production). Since their
initial discovery \citep{cayrel01}, a number of such abundance patterns have been observed, up to now all in extremely metal-poor stars. This could indicate changes in the r-process strength for the same r-process sites.

\subsection{Early Galactic Evolution}
\label{sec:2.2}
\begin{figure}[h]
\includegraphics[scale=.05]{figure2-small}
%
%
\caption{Ratios of [Mg/Fe] (blue uncertainty range, indicating 95\% of observations) and [Eu/Fe] (individual stellar observations shown as 
red error bars) as a function of "metallicity" [Fe/H] for stars in our Galaxy, as displayed in \cite{thielemann15} and taken from a database 
\citep{suda08,suda11}. [X/Y] stands for  $log_{10}$ [(X/Y)/(X/Y)$_\odot$], 
i.e. [Mg/Fe]=0 or [Fe/H]=0 for solar ratios, -1 for 1/10 of solar etc.. Mg shows a relatively flat behavior up to [Fe/H] $\le$ -1, similar to other alpha elements like O, Mg, Si, S, Ar, Ca, Ti, turning down to 
solar values at [Fe/H]=0. This is explained by the early contributions of core-collapse supernovae before type Ia supernovae set in. The real 
scatter is probably smaller than indicated by the blue region, as this is a collection of many observations from different telescopes, 
different observers and different analysis techniques. To the contrary, the scatter of Eu/Fe is larger than two orders of magnitude at low 
metallicities, indicating production sites with a low event rate, and thus taking longer to arrive at average values only in the interval
-2$\le$[Fe/H]$\le$-1.  Such average values are seen for alpha-elements (with core-collapse supernova origin) already in the range -4$\le$[Fe/H]$\le$-3      }
\label{fig_2}       
\end{figure}

As mixing  of ejecta into the interstellar medium is not instantaneous, there will be local inhomogeneities after individual nucleosynthesis events.
Mixing occurs (a) via the plowing of a Sedov-Taylor blast wave through interstellar matter until the (kinetic) explosion energy is used up,
working against the ram pressure of the surrounding medium. For a standard explosion energy of $E=10^{51}$erg (a unit known as 1 Bethe, or 1 foe, an
acronym based on 10 to the fifty-one ergs) and typical densities
of the interstellar medium, this results in mixing with about a few times $10^4$ M$_\odot$. (b) There will be mixing via other more macroscopic phenomena, like e.g. turbulent mixing and/or spiral arm movements with time scales on the order of $10^8$ y, for turbulent mixing being possibly as low as a few  $10^7$ y? While the latter effects can smooth spatial abundance 
gradients, 
the first one will keep the individual composition of a specific explosive event until many other events from different
stellar sources/explosions pollute the interstellar medium in the same vicinity. This causes an integrated average of ejecta 
compositions.  Thus, while we expect an average value of e.g. [Eu/Fe] to occur in late galactic evolution, rare events will lead to large variations at low metallicities, depending on whether  or not a rare nearby strong r-process source polluted the environment.

Neutron star mergers have high predicted ejecta masses of the order of a few times 
$10^{-3}$ to $10^{-2}$ M$_\odot$ of overall r-process matter in dynamical ejecta, and are rare in comparison to regular core-collapse supernovae (CCSNe) with a frequency 
being smaller by a factor of 100 to 1000 \cite{matteucci15,macias16}. Such event rates are also consistent with population 
synthesis studies \cite{chruslinska16} and with (inhomogeneous) chemical evolution calculations 
\citep{argast04,mennekens14,cescutti15,vandervoort15,shen15,wehmeyer15,hirai15,mennekens16}. The latter can follow local 
variations of abundances due the specific contributions by individual explosions. The scatter of r-process elements 
(e.g. Eu) compared to Fe at low metallicities covers more than two orders of magnitude (see Fig.\ref{fig_2}) and indicates production sites with negligible Fe production \cite{cowan05} and a low event rate combined with high ejecta masses in order to explain solar abundances. This causes 
the effect that for [Eu/Fe] the approach to an average ratio occurs only in the interval -2$\le$[Fe/H]$\le$-1. It is shifted in comparison to the behavior of [Mg/Fe], due to the much higher CCSNe rate. The latter permits a much earlier 
approach to an average ratio in the metallicity range [Fe/H]=-3 to -4.

Dwarf galaxies, and especially ultra-faint dwarf galaxies, are polluted only by a few \cite{cohen09,jablonka15,simon15,tsujimoto15} or in extreme cases only one single nucleosynthesis event, as e.g. seen in Reticulum II \cite{ji16a,ji16}. These observations, especially the latter, require events with high r-process ejecta masses, consistent with the
above conclusions from low metallicity stars in the Milky Way. The Milky Way might evolve from an assembly of initially individual dwarf
galaxies where the star formation efficiency and rate can vary in these early components, before the present galaxy emerges
\cite{ishimaru15,hirai15,beniamini16b}.

\subsection{Short Duration Gammy-Ray Bursts and Macronovae (Kilonovae)}
\label{sec:2.3}
While in the previous subsection we discussed overall constraints, i.e. how to reproduce the solar r-process abundance pattern, indications for individual events are harder to obtain. Low metallicity stars can serve here to some extent, as they might only have been polluted by one nucleosynthesis event. A clearer constraint is based on direct observations of a single event, in order to test whether theoretical predictions
for an r-process are underpinned by observational proofs for these objects. Neutron star mergers have been identified with
short duration gamma-ray bursts or macronovae via light curves and spectra of electromagnetic counterparts. These are not yet proofs for a detailed abundance pattern, but the existing observations can only be understood utilizing opacities of (very) heavy elements 
\citep[e.g.][]{metzger10,tanvir13,berger13,grossman14,yang15,fryer15,lippuner15,fernandez16,rosswog17,metzger17}. The radioactive energy emitted from heavy unstable nuclear species, together with its thermalization efficiency, sets the luminosity budget and is therefore crucial for predicting macronova light curves. In modeling the 
macronova accompanying gamma-ray burst 130603B, estimates for the mass ejection could be made \cite{barnes16} in that event. This work also showed that late time macronova light curves can be significantly impacted by alpha-decay from translead isotopes. The latter could actually be a diagnostic test for more detailed ejecta abundances. We want to keep this discussion brief and refer the reader to last year's Annual Review article by Fernandez \& Metzger on electromagnetic signatures of compact binary mergers, which discusses this topic in extended detail \cite{fernandez16}. The message to take home is that only with opacities of very heavy elements the light curves and spectra of these events are explainable, i.e. short durations GRBs produce these heavy elements in sizable amounts. While the observations
integrate over many (also radioactive) elements, a detailed abundance pattern can not be determined, but there is hope to identify specific features with further investigations \cite{barnes16}.

\subsection{Recent Radioactive Additions to the Solar System}
\label{sec:2.4}
While the above discussion points to rare strong r-process events in the early galaxy, there exist other observations, suggesting 
the same in recent history. Long-lived radioactive species can act as witness of recent additions
to the solar system, dependent on their half-lives. For a review on the signature of radioactive isotopes alive in the early solar system see e.g. \citep{davis14}. Two specific isotopes have been utilized in recent years to measure such activities in deep sea 
sediments.
One of them, $^{60}$Fe, has a half-life of $2.6\times 10^6$ y and can indicate recent additions from events occurring up to several million 
years ago. $^{60}$Fe is produced during the evolution and explosion of massive stars (leading to supernovae) \citep{thielemann11,wanajo13}.
It is found in deep-sea sediments which incorporated stellar debris from a nearby explosion about two million years ago 
\citep{knie04,feige15,ludwig15,ludwig16}. Such a contribution is consistent with a supernova origin and related occurrence frequencies, witnessing the last nearby
event. Another isotope utilized, $^{244}$Pu, has a half-life of $8.1\times 10^7$ y and would lead to a collection of quite a number of such supernova 
events. If the strong r-process would take place in every core-collapse supernova from massive stars, 
about 10$^{-4}$-10$^{-5}$ M$_\odot$ of 
r-process matter would need to be ejected per event in order to explain the present day solar abundances. The recent $^{244}$Pu detection \citep{wallner15} 
is lower than expected from such predictions by two orders of magnitude, suggesting that actinide nucleosynthesis is very rare (permitting
substantial decay since the last nearby event) and that 
supernovae did not contribute significantly to it in the solar neighborhood for the past few hundred million years. Thus, in addition to
the inherent problems of (regular) core collapse supernova models to provide conditions required for a strong r-process - also producing the
actinides - these observational constraints from nearby events also challenge regular CCSNe as source of main r-process 
contributions. A recent careful study of the origin of the strong r-process with continuous accretion of interstellar dust grains into the inner solar system \cite{hotokezaka15} concluded that the experimental findings \cite{wallner15} are in agreement with an r-process origin  from a rare event like neutron star mergers. This explains the
$^{244}$Pu existing initially in the very early solar system as well as the low level of more recent additions witnessed in deep-sea sediments over the past few hundred million years.

\section{CONDITIONS FOR MAKING THE HEAVIEST ELEMENTS}
\label{sec:3}

Many sites for the r-process have been suggested in the past, from regular CCSNe, via neutrino-induced processes in outer shells of massive stars, ejecta from compact binary mergers, up to a special class of core collapse supernovae (MHD-jet supernovae) with fast rotation, high magnetic fields and neutron-rich jet ejecta along the poles.
In any of these cases, the production of r-process nuclei occurs in a two stage process, defined by initial explosive burning at high temperatures until charged-particle 
freeze-out during the expansion with a high neutron/seed ratio. This is followed by the rapid capture of neutrons on these seed nuclei, producing the heaviest nuclei. 

\subsection{Explosive Burning and Charged-Particle Freeze-Out}
\label{sec:3.1}
 
In the first stage matter experiences explosive 
burning at high temperatures and is heated to conditions which permit a so-called nuclear-statistical equilibrium (NSE), 
which indicates a full chemical equilibrium of all involved nuclear reactions. At density $\rho$ and temperature $T$ nucleus $i$ - with
neutron number $N_i$, proton number $Z_i$, and mass number $A_i=Z_i+N_i$ - is existing with the abundance $Y_i$, expressed in terms of the abundances of
free neutrons $Y_n$ and protons $Y_p$

\begin{equation}
Y_i=G_i(\rho N_A)^{A_i-1}{A_i^{3/2}\over 2^{A_i}}\left({2\pi\hbar^2
\over m_uk_bT}\right)^{3/2(A_i-1)}\exp(B_i/k_bT)Y_n^{N_i} Y_p^{Z_i},
\label{1:NSE}
\end{equation}
 
where $G_i$ is the partition function of nucleus $i$, $N_A$ is Avogadro's number, $m_u$ the nuclear mass unit, $k_b$ the Boltzmann constant, and $B_i$ the nuclear binding energy of the nucleus.
Beta-decays, electron captures, and neutrino interaction, change the overall proton to nucleon ratio $Y_e=\sum Z_i Y_i /\sum A_i Y_i$
(the denominator is the sum of all mass fractions and therefore equal to unity) and occur on longer time 
scales than particle captures and photodisintegrations. They are not in equilibrium and have to be followed explicitly. Thus, as a function
of time the NSE will follow the corresponding densities $\rho(t)$, temperatures $T(t)$, and $Y_e(t)$, leading to two equations based on total 
mass conservation and the existing $Y_e$

\begin{eqnarray}
\sum_i A_i Y_i&=&Y_n+Y_p+\sum_{i>n,p}
(Z_i+N_i)Y_i(\rho,T,Y_n,Y_p)=1 \\
\sum_i Z_i Y_i&=&Y_p+\sum_{i > p} Z_i Y_i(\rho,T,Y_n,Y_p)=Y_e.
\label{2:NSE}
\end{eqnarray}

In general, very high densities favor large nuclei, due to the high power of
$\rho^{A-1}$, and very high temperatures favor light nuclei, due to 
$(kT)^{-3/2(A-1)}$. In the intermediate regime $\exp (B_i/kT)$ favors tightly bound nuclei with the highest binding
energies in the mass range $A=50-60$ of the Fe-group, but depending on the
given $Y_e$. The width of the composition distribution is determined by the temperature.
Thus, in this first stage of the scenario discussed here, high temperatures cause the (photo-)disintegration of nuclei into neutrons, protons, 
and alpha-particles, due to the energy distribution of the black body photon gas. During the subsequent cooling and expansion of matter, a build-up 
of heavier nuclei sets in, still being governed by the trend of keeping matter in NSE. However, the build-up of nuclei beyond He is hampered
by the need of reaction sequences involving highly unstable $^8$Be (e.g. $\alpha+\alpha+\alpha \rightarrow ^{12}$C or  $\alpha+\alpha+n \rightarrow ^9$Be) 
which are strongly dependent on the density of matter. The first part of these reaction sequences involves a chemical equilibrium for
 $\alpha+\alpha \leftrightarrow ^{8}$Be which is strongly shifted to the left side of the reaction equation, due to the half life of $^{8}$Be 
 ($\tau_{1/2}=6.7 \times 10^{-17}$s). Reasonable amounts of $^{8}$Be, which permit the second stage of these reactions via an alpha or neutron-capture,
 can only be built-up for high densities. The reaction rates for the combined reactions have a quadratic dependence on density in comparison to a linear 
 density dependence in regular fusion reactions. Therefore, for low densities the NSE cannot be kept and after further cooling and freeze-out of 
 charged-particle reactions
an overabundance of alpha particles (helium) remains, permitting only a (much) smaller fraction of heavier elements to be formed than in an NSE
for the intermediate regime (determined by binding energies of nuclei). This result is also called an alpha-rich freeze-out (of charged-particle reactions)
and leads to the fact that (a) the abundance of nuclei heavier than He is (strongly) reduced in comparison to their NSE abundances, and (b) the 
abundance maximum of the (fewer) heavy nuclei is shifted (via final alpha captures) to heavier nuclei in comparison to an NSE. 
While this maximum would normally be around Fe and Ni (the highest binding energies) with A=50-60, it can be shifted up to A about 90.

\begin{figure}[t]
\includegraphics[scale=.085]{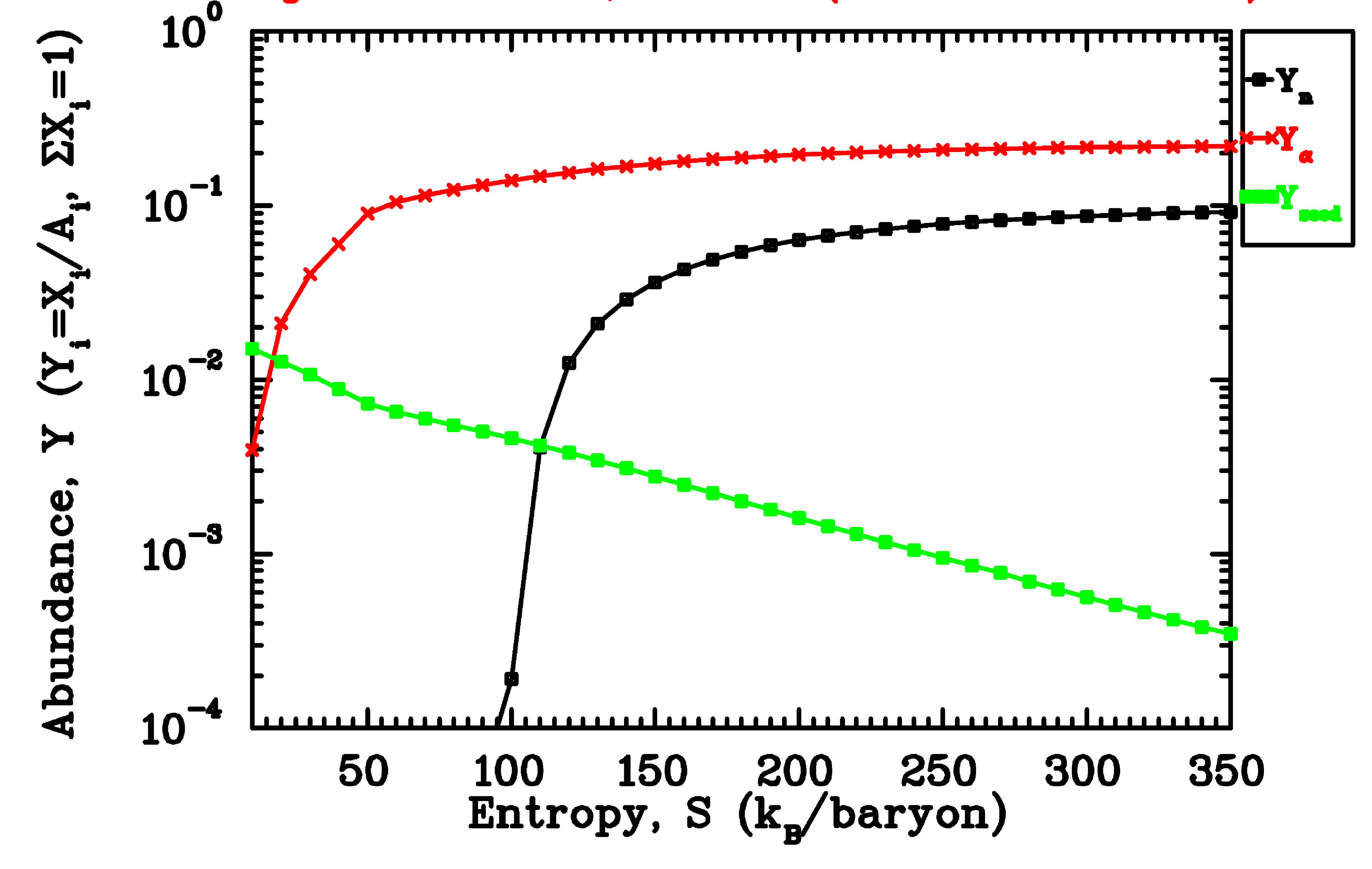}
\includegraphics[scale=.12]{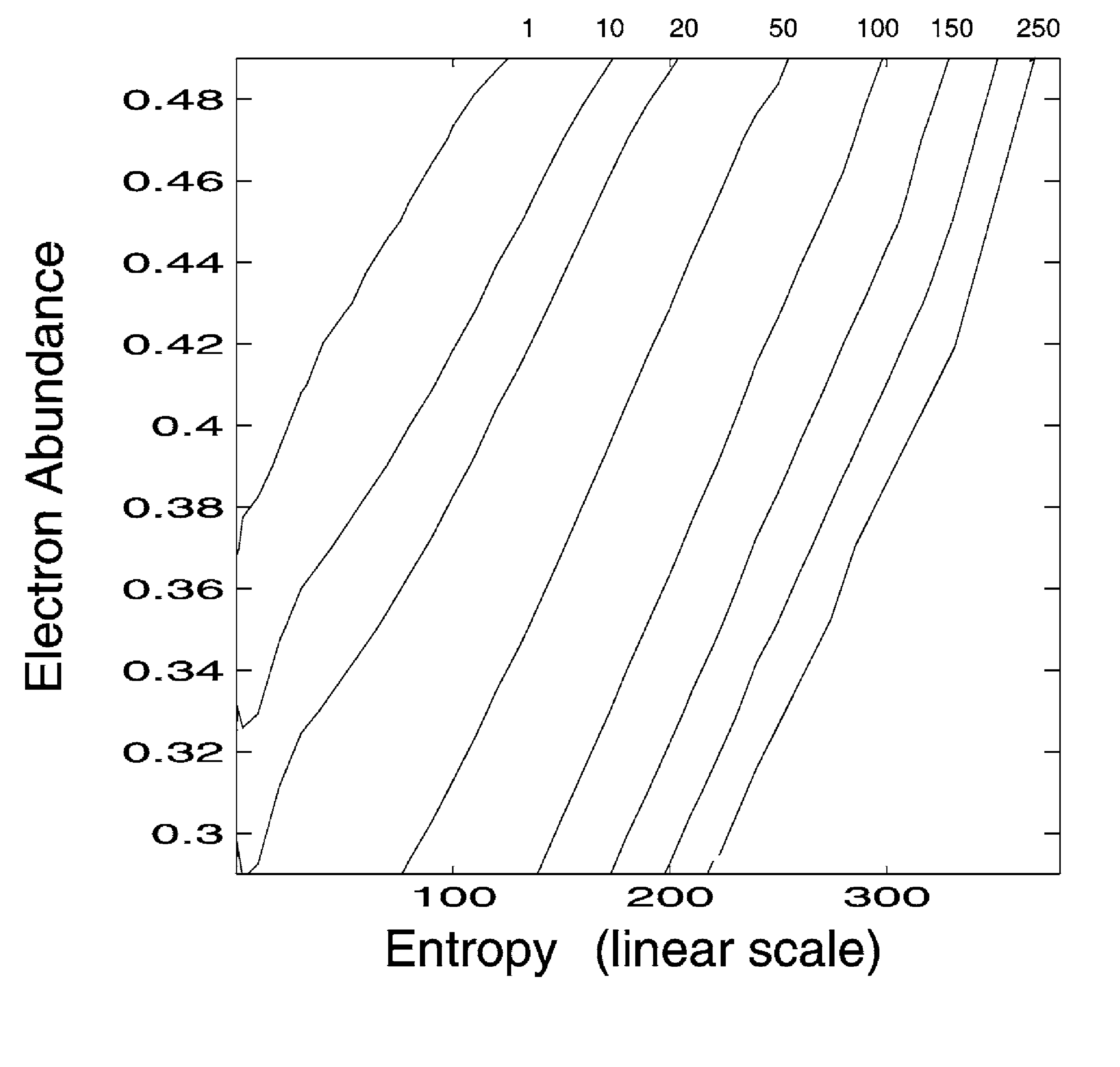}
%
%
\caption{Top: Abundances of neutrons $Y_n$, $^4$He (alpha-particles) $Y_\alpha$, and so-called seed nuclei $Y_{seed}$ (in the mass range 
50$\le$A$\le$100), resulting after the charged particle freeze-out of explosive burning, as a function of entropy in the explosively expanding
plasma, based on results by \cite{farouqi10}. It can be realized that the ratio of neutrons to seed nuclei (n/seed=$Y_n/Y_{seed}$) increases with 
entropy. The number of neutrons per seed nucleus determines whether the heaviest elements (actinides) can be produced in a strong r-process, requiring 
$A_{seed}$+n/seed $\ge$230. 
Bottom: n/seed ratios (shown as contour lines) resulting in expanding hot plasmas from explosive burning as a function of the electron abundance
$Y_e$ and the entropy (measured in $k_b$ per baryon). A strong r-process, producing the actinides with n/seed of 150, requires for moderate 
$Y_e$'s, of about 0.45, entropies beyond 250 \citep{freiburghaus99a}. }
\label{fig_3}       
\end{figure}

In hot environments the total entropy is dominated by the black-body photon gas (radiation) and proportional to $T^3/\rho$ 
\citep{woosley92,roberts10}, i.e. the combination
of high temperatures and low densities leads to high entropies. Thus, high entropies cause an alpha-rich freeze-out, and - dependent on the 
entropy - only small amounts of Fe-group elements are produced, essentially all matter which passed the bottle neck beyond He. This result 
is shown in Fig.\ref{fig_3}a.

The calculation for Fig.\ref{fig_3}a, performed with an expansion time scale equivalent to a free-fall for those conditions and a $Y_e=0.45$, shows how with 
increasing entropies the alpha mass-fraction ($X_\alpha = 4 Y_\alpha$) is approaching unity and the amount of heavier elements (which would provide
the seed nuclei for a later r-process) is going to zero. This is
similar to the big bang, where extremely high entropies permit essentially only elements up to He, and tiny amounts of Li. Opposite to the 
big bang, which is proton-rich, the conditions chosen here ($Y_e=0.45$) are slightly neutron-rich, leading at high entropies predominantly to
He and free neutrons. The small amount of heavier nuclei after this charged-particle freeze-out (in the mass range of A=50-100), depending on
the entropy or alpha-richness of the freeze-out, can then act as seed nuclei for capture of the free neutrons. As prerequisite for an 
r-process, producing nuclei as heavy as the actinides and starting from A=50-100 nuclei, a neutron/seed ratio of about 150 is
required. This ratio is plotted in the form of a contour plot and as a function of entropy and $Y_e$ in Fig.\ref{fig_3}b, based on 
\cite{freiburghaus99a}. 

A different behavior occurs for lower entropies, i.e. the expansions of relatively cold and/or high density matter, as it would exist in ejected neutron star matter. At such low entropies the contour lines for constant n/seed ratios of Fig.\ref{fig_3}b will bend over and become flat, with the resulting neutron/seed ratio being essentially only a function of $Y_e$. In order to obtain then an n/seed ratio of 150, a $Y_e$ of the order 0.1 is required.

%
%

\subsection{Neutron Captures in the r-Process}
\label{sec:3.2}

Once a freeze-out of charged-particle reactions and full chemical equilibrium (NSE) occurrred, resulting in a high neutron/seed ratio, the actual r-process - powered solely by the rapid capture of neutrons - can start, at temperatures below $3 \times 10^9$K and all nuclear reactions have to be followed in full detail. This leads  to three types of terms in 
reaction network equations. The nuclear abundances $Y_i$ enter in this set of equations
and their time derivative can be written in the form

\begin{equation}
{{d Y_i} \over dt}  =  \sum_j P^i _j\ \lambda_j Y_j + \sum_{j,k} P^i _{j,k}\ 
\rho N_A <j,k> Y_j Y_k 
+ \sum_{j,k,l} P^i _{j,k,l}\ \rho^2 N_A^2 <j,k,l> Y_j Y_k Y_l. 
\label{4:network}
\end{equation}

One has to sum over all reaction partners given by the different summation indices. The P's include an integer (positive or negative) factor $N^i$, 
describing whether (and how often) nucleus $i$ is created or destroyed in this reaction, but also correction factors avoiding multiple counting in 
case two or three identical reaction partners are involved. The $\lambda$'s stand
for decay rates (including decays, photodisintegrations, electron captures and neutrino-induced reactions), $<j,k>$ denotes the thermal average for the product of reaction cross section $\sigma$ and relative velocity $v$ of reactions between nuclei $j$ and $k$, while $<j,k,l>$ includes a similar
expression for three-body reactions \citep{nomoto85}. For a survey of computational methods to solve nuclear networks see 
\cite{hixthi99m,timmes99}.
The abundances $Y_i$ are related to number densities $n_i=\rho N_A Y_i$ and mass fractions of the corresponding nuclei via $X_i=A_i Y_i$. Data repositories of experimental and theoretical
reaction rates required as input for equation \ref{4:network} can be found e.g. on the following websites https://groups.nscl.msu.edu/jina/reaclib/db/, 
https://nucastro.org/reaclib.html, and \newline {http://www.kadonis.org/, http://www.astro.ulb.ac.be/pmwiki/Brusslib/HomePage).} A more detailed discussion of modeling nucleosynthesis processes is
given in \cite{thielemann11}.

As charged-particle reactions are frozen at about $3 \times 10^9$K , the only connection between isotopic
chains is given by beta-decays (unless fission will set in, repopulating lighter nuclei from fission fragments). High neutron
densities make the timescales for neutron capture much faster than those for beta-decay and can produce nuclei with neutron separation energies $S_n$ 
of 2 MeV and less.
This is the energy gained (Q-value) when capturing a neutron on nucleus $A-1$ and or the photon energy required to release a neutron from nucleus $A$ via 
photo-disintegration. At the neutron drip-line $S_n$ goes down to 0, i.e. for the high neutron densities of such an r-process it proceeds close to the neutron drip-line. For temperatures around  
$10^9$K, $(\gamma,n)$ photodisintegrations can still be very active for such small reaction
$S_n$-values, as only temperatures related to about $30kT \ge S_n$ are required for these reverse reactions to dominate. With both reaction directions being
faster than process timescales (and beta-decays) a chemical equilibrium can set in between neutron captures and photodisintegrations. In such a case, a 
complete chemical or
nuclear statistical equilibrium (NSE) - discussed in the beginning of this subsection - splits into many (quasi-) equilibrium clusters, representing
each an isotopic chain of heavy nuclei. The abundance distribution in each isotopic chain follows the ratio of two neighboring isotopes

\begin{equation}
{{Y(Z,A+1)}\over {Y(Z,A)}}=n_n {{G(Z,A+1)}\over{2G(Z,A)}}
\Bigl[{{A+1}\over {A}}\Bigr]^{3/2} \Bigl[{{2\pi
\hbar^2}\over {m_uk_bT}}\Bigr]^{3/2} {\rm exp}(S_n(A+1)/k_bT).
\label{1:ngamma} 
\end{equation}

with partition functions $G$ describing the thermal population of excited states, the nuclear-mass unit $m_u$, and
the neutron-separation (or binding) energy of nucleus ($Z,A$+1), $S_n$($A$+1),
being the neutron-capture $Q$-value of nucleus $(Z,A)$.
The abundance ratios are dependent only on $n_n=\rho N_A Y_n$, $T$ and $S_n$. $S_n$ introduces the dependence on nuclear masses, i.e. a nuclear-mass
model for these very neutron-rich unstable nuclei. Under the assumption of an ($n,\gamma$)${\rightleftharpoons}$($\gamma,n$) equilibrium, no
detailed knowledge of neutron-capture cross sections is needed.

\begin{figure}
\includegraphics[scale=0.30]{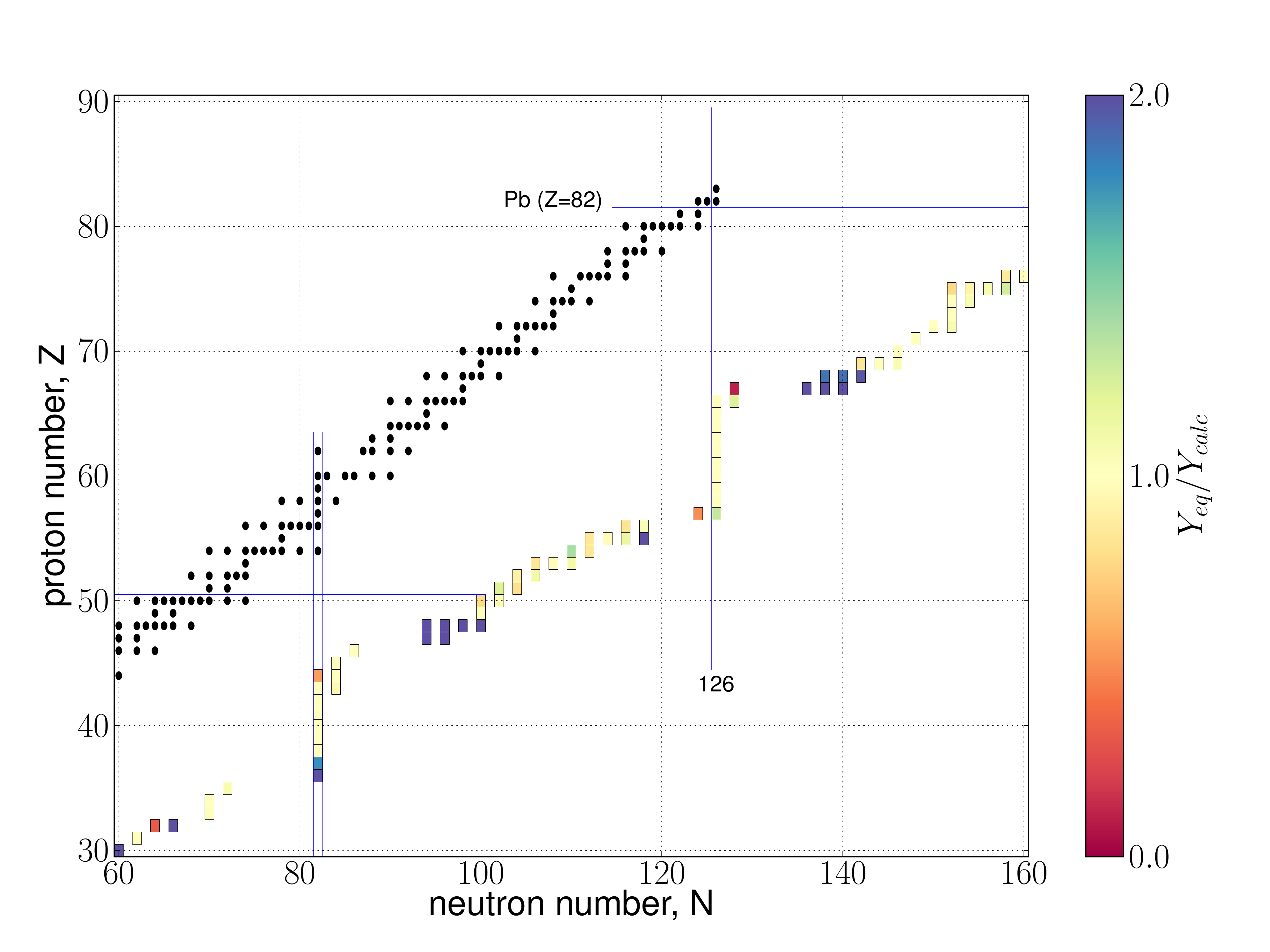}
%
%
\caption{Shown is (a) the line of stability (black squares) and (b) the r-process path, resulting here from a neutron star merger environment
which will be discussed further below \citep{eichler15}. The position of the path follows from a chemical equilibrium between neutron
captures and photo-disintegrations in each isotopic chain ($(n,\gamma)$ - $(\gamma,n)$ equilibrium), determined by the neutron
number density and temperature. However, the calculation was 
performed with a complete nuclear network containing more than 3000 nuclei. The colors along the path indicate how well the full network
calculation follow such an $(n,\gamma)$ - $(\gamma,n)$ equilibrium, It can be seen that such full calculations agree with this equilibrium
approach within a factor of 2 along the r-process path, which continues to the heaviest nuclei. Only in the final phase of the process, when 
neutron number densities and temperatures decline, such an equilibrium freezes out and some final changes of the abundance pattern can occur
due to still continuing neutron captures.   }
\label{fig_4}       
\end{figure}

One fact which can be easily deduced, given that $Y(A+1)/Y(A)$ is first rising with increasing distance from stability, 
close to 1 at the abundance maximum of the isotopic chain, and finally decreasing, is that the abundance maxima in each isotopic chain are
only determined by the neutron number density $n_n$ and the temperature $T$.
Approximating $Y$($Z,A$+1)/$Y(Z,A)$ $\simeq$1 at the maximum and keeping all
other quantities constant, the neutron-separation energy $S_n$ has to be
the same for the abundance maxima in all isotopic chains (see Fig.\ref{fig_4}). It should be said at this point that all present nucleosynthesis calculation are obtained from
full solutions of extended reaction networks determined by the set of equations \ref{4:network}. However, the use of the approximations \ref{1:NSE}, \ref{2:NSE} and \ref{1:ngamma} can act as tests
whether such equilibria exist and aids understanding the numerical results. Fig.\ref{fig_4} displays exactly such a test for the conditions in neutron star mergers \cite{eichler15}.

\begin{figure}
\includegraphics[scale=0.18]{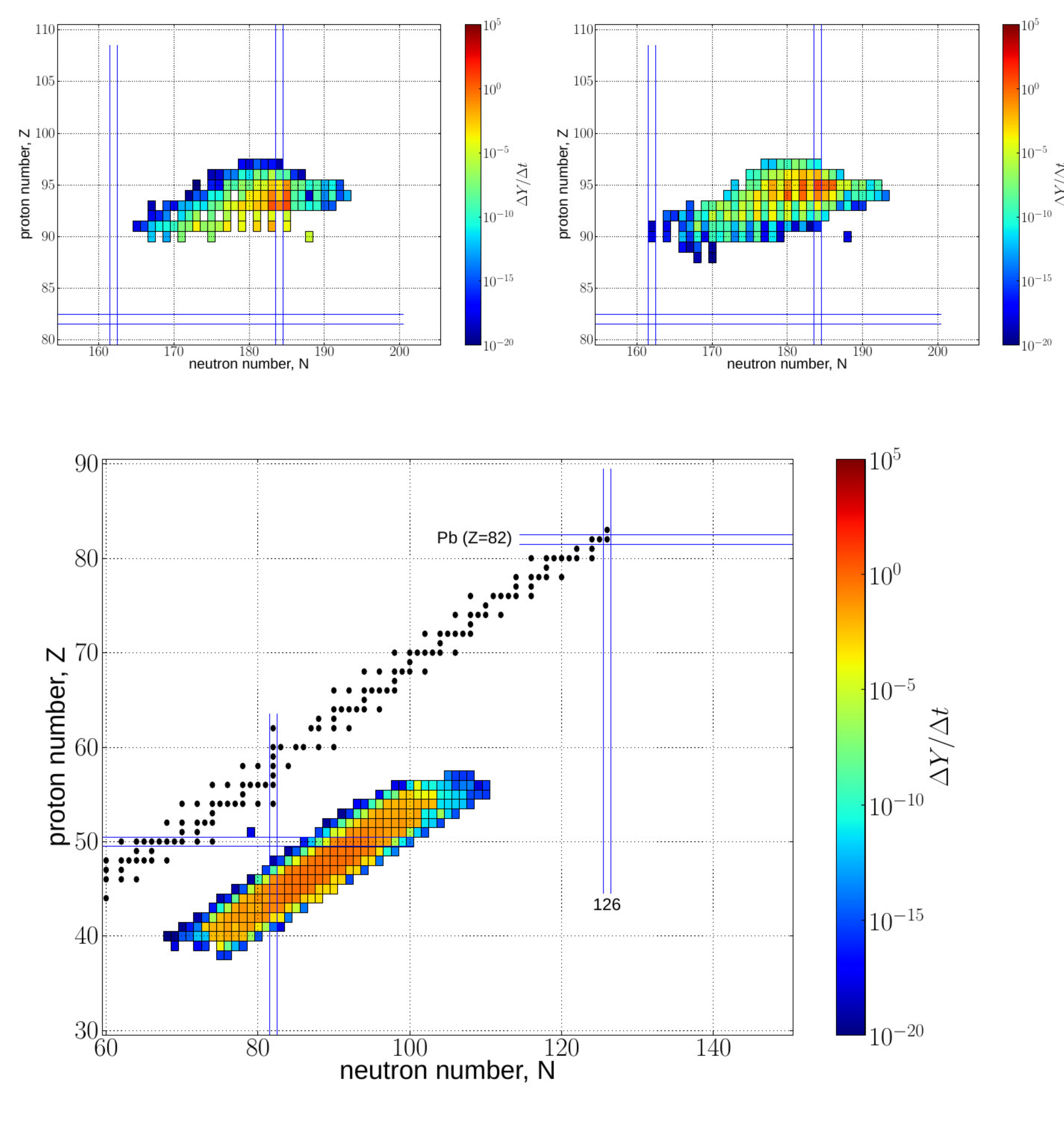}
%
%
\caption{Shown are (color-coded) time derivatives of nuclear abundances Y during an r-process simulation \citep{eichler15}, due to (a) the
destruction via neutron-induced and beta-delayed fission \citep{panov10} and (b) the production of fission fragments \citep{kelic08}. 
The latter are produced in a broad distribution, ranging in mass numbers $A$ from 115 to 155.   }
\label{fig_5}       
\end{figure}

\subsection{The Influence of Nuclear Properties}
\label{sec:3.3}
Fig.\ref{fig_4} shows a contour line of $S_n$ $\simeq$2 MeV for the FRDM mass model \citep{moeller95}.
In addition, it displays the line of stability. As the speed along the r-process path is determined by beta-decays, and they are longest closer 
to stability, abundance maxima will occur at the top end of the kinks in the r-process path at neutron shell closures $N=50, 82, 126$. 
This causes abundance maxima at the appropriate mass numbers $A$ after decay back to stability at the end of the process, which correspond to 
smaller mass numbers $A$ than those for stable nuclei with neutron shell closures.
In environments with sufficiently high neutron densities, the r-process continues to extremely heavy nuclei and finally encounters
the neutron shell closure $N=184$, where fission plays a dominant role. Fig.\ref{fig_5} \citep[based also on simulations by][]{eichler15} shows the 
regions of the nuclear chart where fission dominates and where the fission fragments are located. 

After having discussed here the general working of and the nuclear input for an r-process, the following chapter is related to apply this to neutron star merger 
environments. Independent of the encountered conditions, the influence of nuclear uncertainties should be analyzed, and how they affect the obtained results. Recent tests with respect to mass models, beta-decay half-lives, and fission fragment distributions have been 
performed by \cite{mumpower16,eichler15,goriely13,mendoza15,wu16} utilizing a variety of mass models, beta decay, and fission properties
\cite{moeller95,duflo95,moeller12,goriely13b,wang13,moeller03,marketin15,panov04,panov10,panov16,kelic08,goriely13} as well as analyzing the effect of neutrinos \citep[see e.g.][]{perego14,goriely14,martin15} were the most advanced treatment of neutrino interactions in matter with medium corrections are given by \cite{martinez12,roberts12}. Finally, also tests for neutrino flavor conversion via matter-neutrino resonances have been performed \citep[e.g.][]{zhu16,frensel17}.

\section{r-PROCESS IN COMPACT BINARY MERGERS}

\label{sec:4}

A brief overview with regard to the history of neutron star-black hole or binary neutron star mergers, especially their role
with respect to their nucleosynthesis contributions, has been given in section \ref{sec:1}. Here we want to discuss the main results obtained in recent research \citep[e.g.][]{panov08,goriely11,korobkin12,panov13,bauswein13,goriely13,hotokezaka13,rosswog14,wanajo14,goriely14,perego14,just15,eichler15,martin15,mendoza15,sekiguchi15,ramirez15,hotokezaka15,just16,sekiguchi16,bauswein16,radice16,wu16,roberts17}. 
These include simulations which do not only consider the composition of the dynamical ejecta, but also a neutrino wind composition (along the poles), where matter is ejected from the combined (initially rotationally stabilized) hot neutron star up to the 
point of black hole formation, and afterwards the ejection of matter from (viscous) black hole accretion disks. In the following we discuss these aspects in separate subsections.

\subsection{Dynamical Ejecta}
\label{sec:4.1}

\begin{figure}[]
\includegraphics[scale=.6]{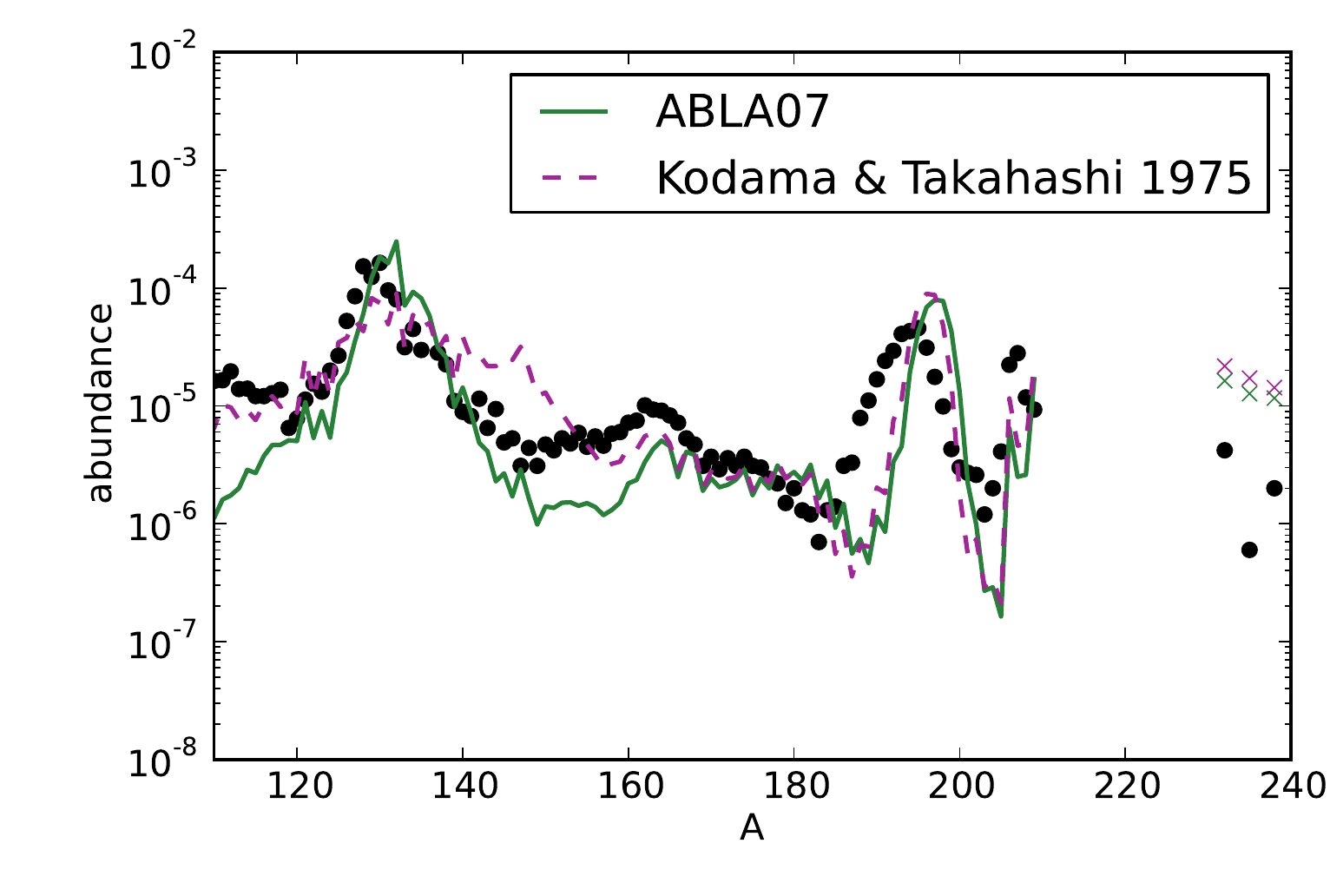}
\includegraphics[scale=.6]{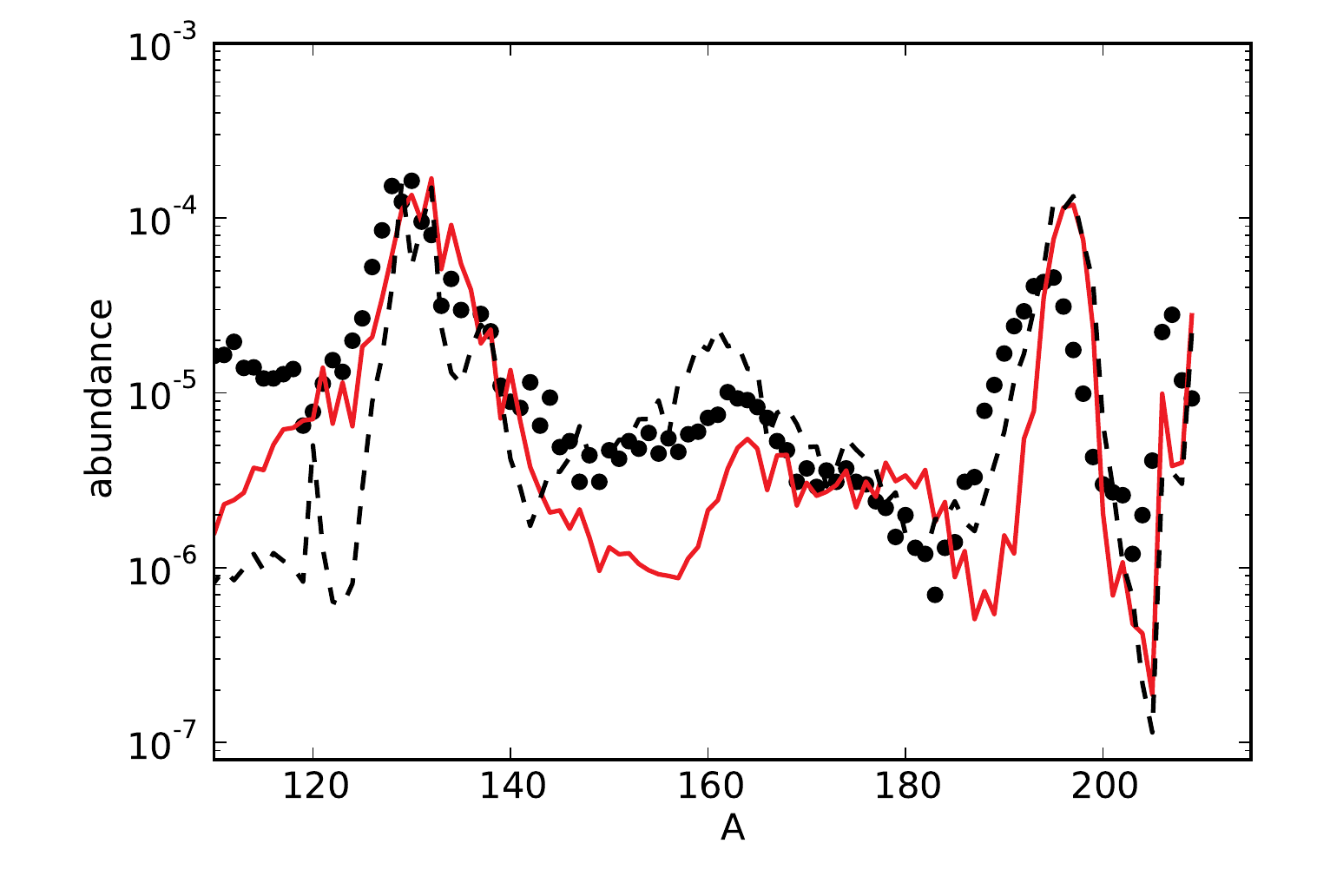}
%
%
\caption{Top: Resulting r-process abundances (in comparison to solar values - black dots) from neutron star merger simulations \citep{eichler15}, making
use of beta-decay half-lives from \citep{moeller03} together with 
a relatively old \citep{kodama75} and a modern set \citep{kelic08} of fragment distributions of fissioning nuclei. However, in both cases 
a shift of the third r-process peak seems to occur in the final phases, driven by neutron capture of the released fission neutrons.  
Bottom: Same as above, but utilizing recent beta-decay half-life predictions \citep{marketin15} (dashed black line) in comparison to the older set (red line, identical to green line from top figure). Faster beta-decays
for heavy nuclei, cause a speed-up of the r-process and deliver (also in the final phases) nuclei which are prone to fission at an earlier time. This way, the late release of fission neutrons occurs earlier, to a large extent before the freeze-out
from ($n,\gamma$)-($\gamma,n$) equilibrium. Therefore, final neutron captures after
freeze-out, which can distort this distribution, are strongly reduced. This can be seen when comparing top and bottom figure. }
\label{fig_6}       
\end{figure}

One of the aspects of earlier investigations, studying only the dynamical ejecta, i.e. matter "thrown out" dynamically after the merger of
two compact objects with very low $Y_e$, was that abundances below the second r-process peak (at A=130) would only result from fission products. 
Thus, lighter r-process elements beyond the Fe-group have already experienced neutron capture and are depleted in the final results. In addition, 
especially for Newtonian calculations, material had the tendency to be possibly too neutron-rich.
This led to large amounts of very heavy nuclei prone to fission, remaining close to the end of the simulations. While initial conditions during
the working of the r-process seem perfect to reproduce the second and third r-process peak and their positions (see the location of the kinks in the 
contour line of constant neutron separation energy in Fig.\ref{fig_4}), during the final phase the fission of 
the heaviest nuclei produces large amounts of neutrons. If this happens during/after the freeze-out from $(n,\gamma)$- ($\gamma,n)$ equilibrium, 
these neutrons can modify the overall abundance pattern inherited from the earlier equilibrium, especially shifting the third
r-process peak. A number of tests, based on latest knowledge of nuclear physics far from stability, have been performed and can improve the
overall abundance pattern. This relates to mass model properties like fission probabilities and fragment distributions (see Fig.\ref{fig_6}) as 
well as beta-decay half-lives \citep{marketin15,panov16}, which speed up the production of the heaviest nuclei and lead to the fact that the final 
phase of fission sets in earlier with respect to the freeze-out and the smaller release of late neutrons has less effect on the pattern of the 
third r-process peak \citep[see Fig.\ref{fig_6}ab and][]{eichler15}.

%
%

\subsection{Neutrino Winds and the Effect of Neutrino Spectra}
\label{sec:4.2}
Another aspect is that also a "neutrino-wind" (similar to that in CCSNe) from the hot, very massive combined object of the two
neutron stars will contribute to the nucleosynthesis of these events after the dynamic ejecta discussed above. This hot central object, supported by high temperatures and rotation, will not collapse to a black hole immediately, and surrounding matter experiences the 
radiation of neutrinos and antineutrinos, changing the $Y_e$ by the reactions

\begin{eqnarray}
\nu_e + n & \rightarrow &p+ e^- \\
\bar\nu_e + p& \rightarrow &n+ e^+ .
\end{eqnarray}

\noindent These reactions turn matter only neutron-rich if the average antineutrino energy $\langle\epsilon_{\bar\nu_e}\rangle$ is higher than the average neutrino
energy $\langle\epsilon_{\nu_e}\rangle$ by 4 times the neutron-proton mass difference $\Delta$ for similar (electron) neutrino $L_{\nu_e}$ and antineutrino $L_{\bar\nu_e}$ luminosities. This was pointed out initially in  \citep{qian96}, leading - when approaching equilibrium conditions for neutrino and antineutrino captures - to 

\begin{equation}
Y_e={\Bigl[ 1 + \frac{{L_{\bar\nu_e} (\langle\epsilon_{\bar\nu_e}\rangle - 2\Delta + 1.2\Delta^2/\langle\epsilon_{\bar\nu_e}\rangle)}}{{L_{\nu_e} ( \langle\epsilon_{\nu_e}\rangle + 2\Delta + 1.2\Delta^2/\langle\epsilon_{\nu_e}\rangle)}} \Bigr]}^{-1}
\end{equation}

\begin{figure}[h]
\includegraphics[scale=.4]{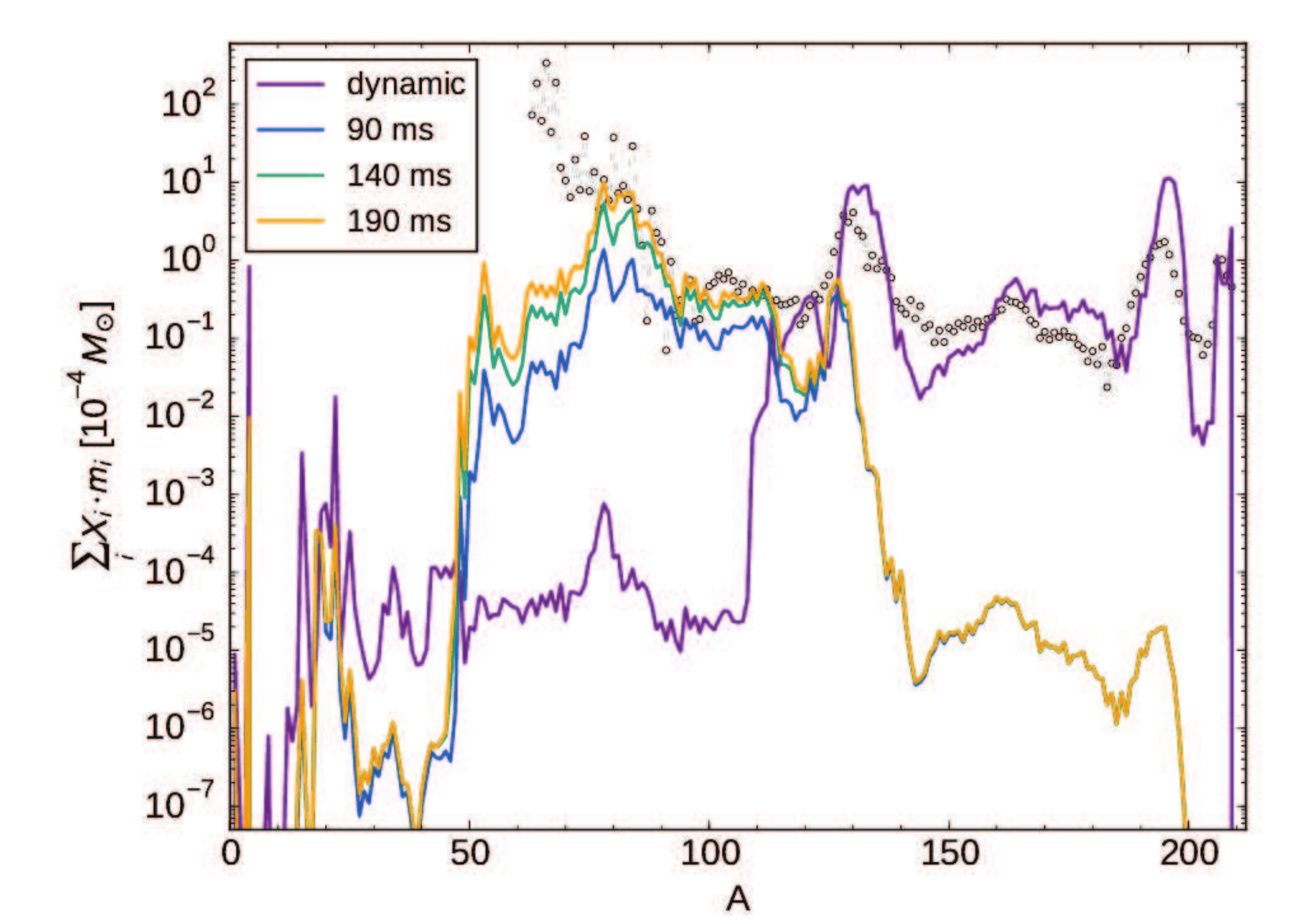}
%
%
\caption{Neutrino wind contribution to neutron star merger ejecta, dependent on the delay time between the merger and BH formation \cite{martin15}.
In comparison also the dynamic ejecta of \cite{korobkin12} are shown. The neutrino wind, ejected dominantly in polar regions, contributes
nuclei with $A<130$, due to the effect of the neutrinos on $Y_e$.  }
\label{fig_7}       
\end{figure}

\noindent For further details and in-medium corrections for neutrons and protons in comparison to their treatment as free particles see \cite{martinez12,roberts12}. Thus, in most cases the energetically favorable first reaction wins, changing $Y_e$ from
the initial (neutron-rich) conditions towards values closer to $Y_e=0.5$, which leads only to a weak r-process and produces matter below the
second r-process peak. A first estimate of this outcome was presented in \cite{rosswog14}. More detailed results have been shown in
\cite{perego14,martin15,radice16,lehner16,roberts17}, see e.g. Fig.\ref{fig_7}.

There exists a related aspect, which also affects the dynamical ejecta. A number of simulations discussed above
have been performed with Newtonian physics, i.e. non-relativistically. For neutron stars, and especially finally resulting black holes
the role of general relativity is important and leads to deeper gravitational potentials plus higher
temperatures experienced by the matter involved. This increases the importance of electron-positron pairs, positron captures on neutrons
and also the effect of neutrino radiation even for the dynamical ejecta. In total this increases $Y_e$ from 0.05 or less in pure neutron star
matter to values around 0.1-0.15 \citep{wanajo14,goriely14,sekiguchi15,sekiguchi16} for dynamical ejecta and to even higher values in the neutrino wind. As a result less fission cycling occurs, which produces less late emission of fission 
neutrons, and therefore avoids some of the deficiencies of the abundance patterns discussed above with respect to
Fig.\ref{fig_6}, also seen in the dynamic ejecta component of Fig.\ref{fig_7}.

Possible changes of $Y_e$ can also be attained by the modification of neutrino and antineutrino spectra due to neutrino flavor conversion. There have
been a number of tests to verify such neutrino conversions via matter-neutrino resonances \cite{malkus12,foucart15,malkus16,zhu16,frensel17}. The more complicated geometry of
a disk environment in comparison to CCSNe permits until presently only single-angle approximations which might limit the accuracy of present results.
But the existing investigations clearly point to the potential that $Y_e$, and thus the resulting nucleosynthesis, can be affected.

\begin{figure}[h]
\includegraphics[scale=.8]{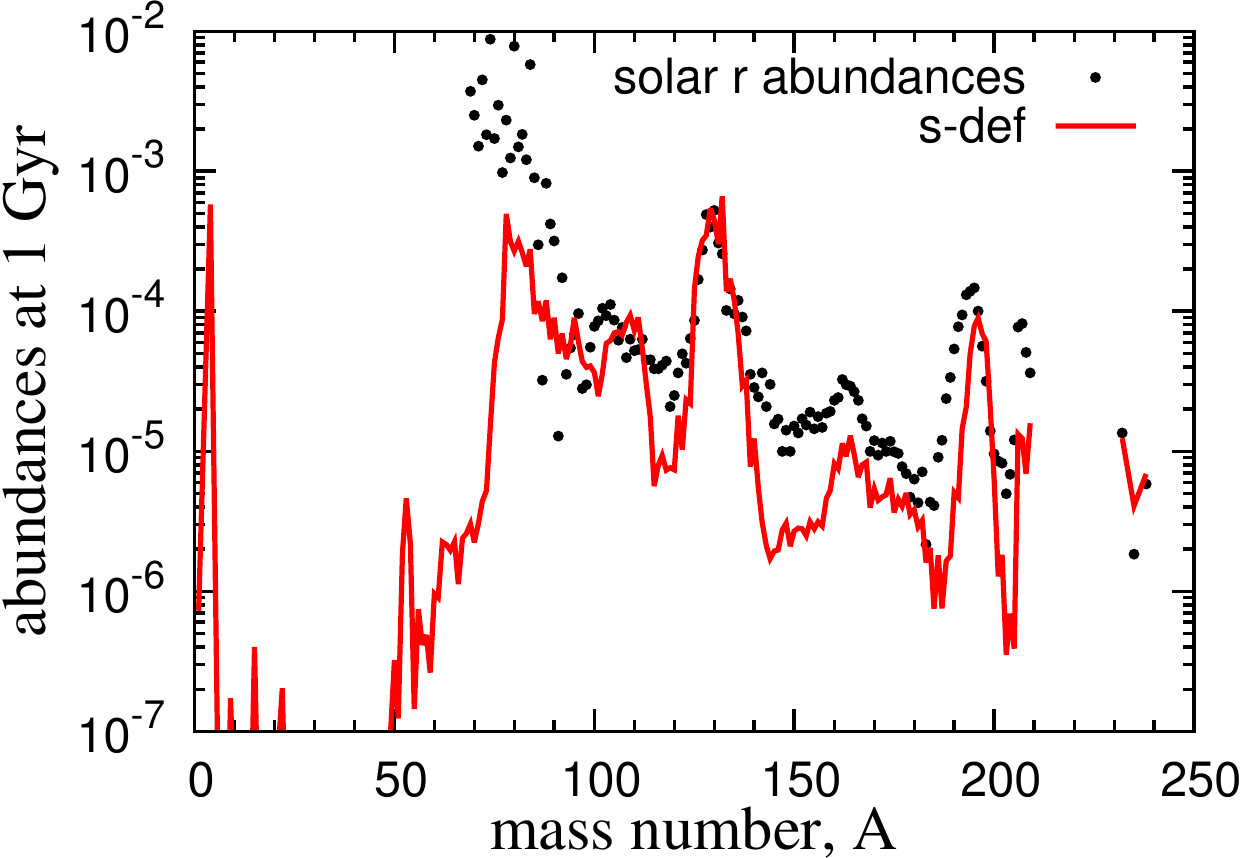}
%
%
\caption{Resulting r-process abundances (in comparison to solar values - black dots) from black hole accretion disk simulations \citep{wu16}, making
use of a black hole mass of 3 M$_\odot$, a disk mass of 0.03 M$_\odot$, an initial $Y_e$ of 0.1, entropy per baryon of 8$k_b$, an alpha parameter of the viscous disk of 0.03, and a vanishing
black hole spin.   }
\label{fig_8}       
\end{figure}

\subsection{Black Hole Accretion Disks}
\label{sec:4.3}

After the stabilizing effect of rotation of the merged quite massive object fades, in most cases (being beyond the maximum mass of cold neutron stars) a central black hole will form. Such environments, resulting
as the final fate of neutron star mergers, require investigations into disk winds from black hole accretion disks, which had initially been tested as sites of heavy element nucleosynthesis
\citep{surman06,surman08,surman14}. Detailed simulations for these sites resulting from binary compact object mergers have been performed in recent years \cite{just15,fernandez16,just16,wu16},
leading to predictions of comparable masses in dynamical ejecta and disk outflows (with a slight dominance of dynamical ejecta for neutron star mergers and the opposite effect for neutron star - black
hole mergers \cite{fernandez16}). 
Latest results for disk outflows \cite{wu16} are displayed in Fig.\ref{fig_8}, which shows the integrated ejecta of all tracer particles. 
This underlines that outflows alone can
produce a robust abundance pattern around the second r-process peak at  A = 130, with a significant production of A $\le$ 130 nuclei. Disc outflows also reach the third peak
at A = 195 in most of their simulations. The detailed results depend on the disk viscosity, initial mass or entropy of the torus, the black hole spin, and (of course) the nuclear physics input. Especially the production of heavy (A=195) nuclei is affected by the uncertainties of these disk properties. However, such a deficit can be easily counterbalanced by the dynamic ejecta, as the
total nucleosynthesis of the merger includes the components of the dynamic ejecta, the neutrino wind, and the BH accretion disk.

\section{A NEED FOR AN r-PROCESS CONTRIBUTION FROM MASSIVE SINGLE STARS?}
\label{sec:5}
\begin{figure}[t]

\includegraphics[scale=0.25]{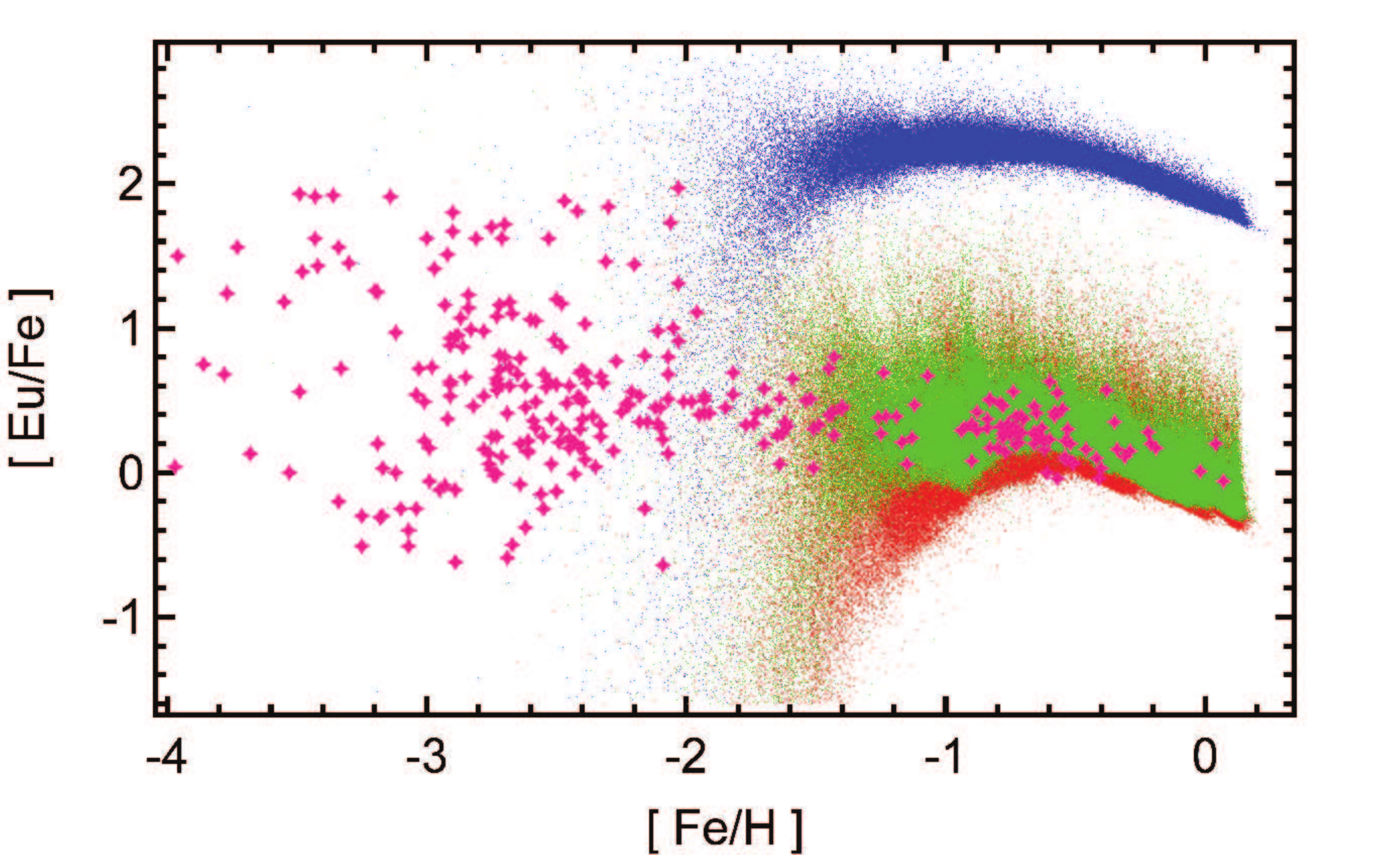}
%
%
\caption{Influence of coalescence time scale and neutron star merger probability on Eu-abundances in galactic chemical evolution. 
Magenta stars represent observations. Red dots correspond to model star abundances as in \cite{argast04}. The coalescence time scale utilized 
is $10^8$ years with a typical probability consistent with population synthesis \citep{wehmeyer15}. Green
dots illustrate the effect on the abundances if the coalescence time scale is shorter (around $10^6$ years). Blue dots show
the abundance change if the probability of neutron star mergers is increased. Within this treatment of
galactic chemical evolution, none of these options would permit a fit with observations of low metallicity stars in the metallicity range
-4$\le$[Fe/H]$\le$-2.5.
     }
\label{fig_9}       
\end{figure}

The previous sections showed that compact binary mergers can reproduce in all cases the heavy (if not most of the) solar r-process abundances, they can explain short duration GRBs and related macronova(kilonova) events, they are rare, consistent with low metallicity observations and deep sea sediments, and in combination of ejecta masses and occurrence frequency they can also explain the total amount of solar r-process matter (within the given uncertainties). There seems to be only one caveat. A binary neutron star 
merger requires two prior supernova events (producing the two neutron stars and e.g. Fe-ejecta) plus the gravitational wave inspiral 
leading to the merger. There is a time delay between the Fe-producing supernovae and r-process ejection which can shift the appearance of a typical r-process tracer like Eu to higher metallicities [Fe/H] (see Fig.\ref{fig_9}), as discussed in \cite{argast04,cescutti15,wehmeyer15}. Such results rely to some extent on the coalescence times (and their distribution) in binary systems, the local star formation rate, and
the amount of mixing of the ejecta with the surrounding interstellar medium, \citep[see also][]{ramirez15}. The results shown in Fig.\ref{fig_9} are based on mixing with the surrounding medium via a Sedov-Taylor
blast wave, i.e. typically of the order $5\times 10^{-4}$M$_\odot$ and with varying coalescense times. The latter seems not to solve the problem to reproduce the [Eu/Fe] ratios in low metallicity
stars.

However, there exist other galactic mixing events on varying timescales (like turbulent mixing) which could blur the picture.
Relatively low resolutions in global galaxy evolution models with smooth particle 
hydrodynamics simulations \citep{vandervoort15,shen15} can wash out the behavior of Fig.(\ref{fig_9}) at low metallicities, but a high resolution run in \citep{vandervoort15} recovers it (see their Fig.4).
The history of the local star formation rate can differ, if the Galaxy formed from small substructures which merge at late times in galactic evolution \citep{ishimaru15,hirai15}. Such aspects still need
to be worked out.
Alternatively a rare class of CCSNe, exploding earlier in galactic evolution with negligible time delay to star formation, could contribute at low metallicities with negligible time delay to star formation. Early suggestions that so-called electron-capture supernovae in the stellar mass range 8-10 M$_\odot$  \citep{kitaura06,janka08,wanajo09,wanajo11} would be able to
produce a strong r-process were never confirmed, and they would also not correspond to rare events. However, other objects driven by strong magnetic fields and fast rotation (possibly about 1\% or less of all core-collapse supernovae), leaving behind 10$^{15}$ Gauss neutron stars (magnetars), might play a significant role. Such magneto-rotational SNe show similar characteristics in the 
amount of r-process ejecta and possibly the occurrence frequency as neutron star mergers, but - because these objects result from massive single stars - they 
do not experience the delay of binary evolution \citep{fujimoto08,ono12,winteler12,moesta14,nishimura15,moesta15,nishimura16}. This might be interesting with respect to the subdivision of short duration GRBs in those with a delay and those following directly the star formation rate \cite{wanderman15}.

Such magneto-rotational SNe, being also rare events and prolific in r-process ejecta, could enter galactic evolution at lowest metallicities with a similar scatter as binary compact mergers. Existing observations show evidence for the occurrence of MHD-jet supernovae \citep[magnetars][]{greiner15}. In inhomogeneous galactic evolution simulations without extended turbulent mixing a superposition of MHD-jet supernovae and neutron star mergers can match observations 
from lowest metallicities up to present. We should keep in mind that there exist uncertainties in mixing processes, star formation rates etc. which will affect the 
behavior at lowest metallicities.
However, as shown in \cite{nishimura15,nishimura16}, dependent on rotation frequency, magnetic fields, and the impact of 
neutrino heating in comparison to the strength of magnetic fields, the strength of the r-process can vary, while neutron star 
mergers seem to predict a robust and unchangeable abundance pattern. At low
metallicities, there exist observations with a somewhat changing Eu/U ratio, indicating to which extent the production of actinides is robustly
coupled to Eu. Few events with a regular r-process pattern but changing amounts of actinides are all seen at metallicities around
[Fe/H]=-3. Thus, such variations, not expected from compact binary mergers might point to the effect of MHD supernovae at low metallicities.
It is reasonable to expect that at that low metallicity MHD SNe are more frequent than in the present Galaxy. Low metallicity stars
have smaller amounts of wind/mass and (therefore) angular momentum loss, providing more promising initial conditions at the
onset of collapse for these events.

\section{CONCLUSIONS}
\label{sec:6}
This review summarizes our present knowledge of r-process conditions in compact binary mergers, their ability to produce a solar abundance pattern, their role in galactic evolution and recent additions to the solar system, and finally also some open questions which still need to be solved or complemented by other sites:
\begin{enumerate}
\item It has been shown in extended sets of simulations that compact binary mergers are prolific sites of r-process nucleosynthesis, leading to 
about a few times 10$^{-3}$ to 10$^{-2}$ M$_\odot$ of ejected r-process matter in the dynamical ejecta and possibly a similar amount via black hole accretion disk wind (see Figure 4 in \cite{fernandez16} and Figure 2 in \cite{rosswog17}). When including
all components - from dynamic ejecta over neutrino winds and final, viscous black hole accretions disks - they produce not only the heaviest r-process nuclei but also significant
amounts of the standard solar r-process abundances for mass numbers with A$<$130. The sizable production of r-process matter requires that these events are rare if they are responsible for 
reproducing all of galactic r-process material. 
\item Radioactive tracers like $^{244}$Pu as well as $^{60}$Fe are found in deep sea sediments. The production of $^{60}$Fe in frequent 
events, related to regular CCSNe and/or electron capture supernovae, is supported by the latest contribution dating back about 2 million years. On the contrary, the amount of $^{244}$Pu found
in these sediments is lower than expected by about a factor of 100, if a quasi-continuous production is assumed. This points to substantial decay since the last addition and to much rarer events.
\item Observations of lowest metallicity stars in our Galaxy and (ultra-faint) dwarf galaxies show substantial "pollution" by r-process elements, indicating a production site with a low event
rate and consistent high amount of r-process ejecta in order to explain solar abundances. This is also underlined by the large 
scatter of Eu/Fe (Eu being an r-process element and Fe stemming from CCSNe at these low metallicities) seen in the earliest stars of the Galaxy, indicating that in a not yet well mixed 
interstellar medium the products of regular CCSNe and these rare events vary substantially.
\item We also know that neutron star mergers (or neutron-star black hole mergers) are related to short-duration gamma-ray bursts and electromagnetic
counterparts (macronovae). The latter can only be explained if the opacity of ejected matter is dominated by heavy elements. Population synthesis supports that these events are very rare 
(probably about 1/100 of the CCSN frequency).
\item The major open question is whether products of the neutron star merger r-process can explain the observations of r-process elements seen 
already at metallicities of [Fe/H]$\le$ -3. As the supernovae which produce the neutron stars of a merger already lead to a substantial floor of Fe,
i.e. enhance [Fe/H], only substantial turbulent mixing of interstellar medium matter in the early Galaxy could reproduce these observations in galactic chemical evolution calculations.
\item There exist observational indications of 10$^{15}$ Gauss neutron stars. 
A rare class of CCSNe driven by a magneto-rotational mechanism could lead to such neutron stars with immense magnetic fields and 
produce r-process matter ejected in polar jets. However, predictions from stellar evolution about the distribution of 
magnetic fields and rotation rates before core collapse are needed in order to understand the initial conditions possibly leading to such events and 
the role of the magneto-rotational instability (MRI) during the collapse/explosion phase has to be investigated.
\item Such objects, very likely also with a low event rate of the order 1/100 of regular CCSNe could possibly avoid the problems of the neutron star merger scenario at low metallicities, as they are related to massive single stars and do not experience 
any delay in comparison to regular CCSNe.
\item Independent of these points related to astrophysical observations and modelling complex astrophysical sites, the final test whether the detailed abundance pattern of heavy elements can 
be reproduced relies on a deep knowledge and understanding of nuclear properties which enter such calculations, from masses far from stability over weak interactions, determining beta-decay properties, electron/positron captures, and neutrino properties and interaction with matter, up to fission barriers and 
fission fragment distributions. And in addition, the equation of state utilized at highest densities and temperatures sets the conditions for such environments.
\end{enumerate}

\section*{ACKNOWLEDGMENTS}
We want to thank all our collaborators with whom we investigated r-process sites and conditions, as well as everybody with whom
we had enlightening discussions. These include Almudena Arcones, Shawn Bishop, Gabriele Cescutti, Cristina Chiappini, John Cowan, Khalil Farouqi, Brad Gibson, Maik Frensel,
Chris Fryer, Samuel Giuliani,
Yuhri Ishimaru, Thomas Janka, Oliver Just, Roger K\"appeli, Oleg Korobkin, Karl-Ludwig Kratz, Karlheinz Langanke, Jim Lattimer, Matthias Liebend\"orfer, Andreas Lohs, Lucio Mayer, Gail McLaughlin, Tomislav Marketin, Dirk Martin, Gabriel 
Martinez-Pinedo, Francesca Matteucci, Nobuya Nishimura, Francesco Pannarale,
Albino Perego, Marco Pignatari, Tsvi Piran, David Radice, Thomas Rauscher, Stephan Rosswog, Chris Sneden, Rebecca Surman, Tomoya Takiwaki, Masaomi Tanaka,
Cristina Volpe, Shinja Wanajo, Christian Winteler, Meng-Ru Wu and many others.
This research was supported by the Swiss SNF (via a regular research grant and a SCOPES project), an ERC Advanced
grant from the European Commission (FISH), and the Russian Science Foundation (project No. 16-12-10519).

%
\bibliography{references}
\bibliographystyle{ar-style5}

\end{document}